\documentclass[english,11pt]{article}
\pdfoutput=1
\usepackage{epsf,amsmath,amssymb,graphicx,dcolumn}
\usepackage{scalefnt,subfigure,ulem}
\usepackage{cite,hyperref}
\usepackage{todonotes}
\usepackage{color}
\usepackage{rotating}
\definecolor{lightblue}{rgb}{.7,.8,1}
\def\bal#1\eal{\begin{align}#1\end{align}}

\newcommand{\pdf}{{\abbrev PDF}}
\newcommand{\qcd}{{\abbrev QCD}}

\newcommand{\abbrev}{\scalefont{.9}}

\newcommand{\mbottom}{m_b}

\newcommand{\fig}[1]{Fig.\,\ref{#1}}

\newcommand{\sct}[1]{Sect.\,\ref{#1}}

\newcommand{\lhc}{{\abbrev LHC}}
\newcommand{\sm}{{\abbrev SM}}

\newcommand{\lo}{\text{\abbrev LO}}
\newcommand{\nlo}{\text{\abbrev NLO}}
\newcommand{\nnlo}{\text{\abbrev NNLO}}
\newcommand{\llog}{\text{\abbrev LL}}
\newcommand{\nll}{\text{\abbrev NLL}}
\newcommand{\nnll}{\text{\abbrev NNLL}}

\newcommand{\mstw}{{\abbrev MSTW2008}}

\newcommand{\reference}[1]{Ref.\,\cite{#1}}

\title{\vspace*{-6em}
  \begin{flushright}
    {\sf\small
      WUB/12-23 
    }
  \end{flushright}
\vspace*{2em}
Top- and bottom-mass effects in hadronic Higgs production at small transverse momenta through \lo{}+\nll{}}
\author{Hendrik Mantler, Marius Wiesemann\\[2em]
{\it Fachbereich C,
  Bergische Universit\"at Wuppertal,}\\[-.3em] {\it 42097 Wuppertal,
  Germany}\\
{\small\tt hendrik.mantler@uni-wuppertal.de}\\[-.3em]
{\small\tt m.wiesemann@uni-wuppertal.de}
}

\date{}
\begin{document}
\maketitle

\begin{abstract}
\noindent
The resummed transverse momentum distribution of the Higgs boson in gluon fusion through \lo{}+\nll{} for small transverse momenta is considered, where the Higgs is produced through a top- and bottom-quark loop. We study the mass effects with respect to the infinite top-mass approach. The top-mass effects are small and the heavy-top limit is valid to better than $4\%$ as long as the Higgs' transverse momentum stays below $150$\,GeV. When the bottom loop is considered as well, the discrepancy reaches up to about $10\%$. We conclude that bottom-mass effects  cannot be included in a reasonable manner by a naive reweighting procedure in the heavy-top limit. We compare our results to an earlier, alternative approach based on {\tt POWHEG}.
\end{abstract}

\section{Introduction}
The recent discovery of a scalar particle \cite{Collaboration2012,ATLAS2012}, which may turn out to be the standard model (\sm{}) Higgs boson, is one of the biggest achievement of particle physics in the last years. The next task is to verify that this particle breaks the electroweak symmetry in a certain theory, e.g. the \sm{}, so that massive particles obtain their masses through the Higgs mechanism. For this purpose its fermionic and bosonic couplings need to be determined, using precision predictions from the theoretical side \cite{Dittmaier2011} and comparing them to experimental data. Besides the total cross section it might be helpful to use differential quantities to disentangle the Higgs couplings to the various particles. 

The most important production mechanism of the Higgs boson in the \sm{} is the gluon fusion process, where the Higgs-gluon coupling is mediated through a top-quark loop. Higher order corrections are usually calculated in the so-called heavy-top limit, which is an effective theory approach, where the top quark is assumed to be infinitely heavy. Calculations that keep track of the full top-mass dependence can usually be performed only at one perturbative order lower than in the heavy-top approximation. The uncertainty constituted by this approach is very specific to hadronic Higgs production in the \sm{}. It has been shown for the total inclusive cross section at next-to-next-to-leading order (\nnlo{}) \cite{Harlander2002,Anastasiou2002,Ravindran2003} that for a Higgs mass lower than the top mass the heavy-top approximation is valid to better than $1\%$ \cite{Marzani2008,Harlander2009,Harlander2010,Pak2009}. Rather few studies aim to validate the heavy-top limit for distributions \cite{Harlander2012,DelDuca2001a,Alwall2012,Bagnaschi2012}. % Until recently \cite{Harlander2012} none of those studies exceeded the leading order \cite{DelDuca2001a,Alwall2012,Bagnaschi2012}.
Generally speaking, it was found that the heavy-top limit works well as long as the transverse momentum of the Higgs is below the top mass.

Another specific uncertainty of hadronic Higgs production in the \sm{} emerges from the bottom-loop contribution\footnote{Throughout this paper we consider the interference terms of the top- and bottom-quark amplitudes as part of the bottom-quark contribution.}. Although suppressed by their couplings, in gluon fusion the Higgs-gluon coupling can be mediated by any quark loop. While the bottom quark has a sizable contribution to hadronic Higgs production, the four lightest quarks $q\in\{u,d,s,c\}$ are usually omitted. There is no effective theory approach for the bottom quark feasible with current technology. Thus for the bottom-quark contribution one has to stick to the perturbative order where the bottom loop can be included. Besides that, one approach which might be followed is to account not only for top- but also for bottom-mass effects by including higher order corrections through reweighting of the mass effects at lower order by the heavy-top limit. We will argue that for certain quantities this appears not to be a good approximation.

Similar to the top-quark effects, rather few studies quantify the importance of the $b$-loop contribution in the \sm{} \cite{Harlander:2003,Alwall2012,Bagnaschi2012} and explore the best way to include it \cite{deFlorian:2009,Anastasiou:2010,Baglio2010}. In the case of the total cross section, for example, it contributes $7\%$ at next-to-leading order (\nlo{}) for a Higgs mass of $125$\,GeV at the Large Hadron Collider (\lhc{}) with $8$ TeV collider energy \cite{Spira:1995,Harlander:2003} and cannot be approximated by the heavy-top limit. A similar study to the one in this paper has been done in \reference{Bagnaschi2012}, where the matched parton shower in the {\tt POWHEG} framework~\cite{Alioli:2008} with full top- and bottom-mass dependence is studied.

One of the most important differential observables is the transverse momentum distribution of the Higgs boson. In the effective theory approach of the gluon fusion process it has been calculated numerically \cite{deFlorian:1999,Ravindran:2002dc} and analytically \cite{Glosser:2002} at \nlo{}. Furthermore, the fully differential Higgs cross section is known through \nnlo{} \cite{Anastasiou2004,Catani2007}. These calculations are only valid for sufficiently high transverse momenta, because they diverge logarithmically in the limit where the transverse momentum of the Higgs vanishes. To obtain a reliable prediction in the small transverse momentum region, these logarithmically enhanced terms need to be resummed to all orders in perturbation theory. This has been worked out in the heavy-top limit at leading logarithmic (\llog{}), next-to-leading logarithmic (\nll{}) \cite{catani:1988,Yuan:1991,kauffman:1991} and next-to-next-to-leading logarithmic accuracy (\nnll{}) \cite{Bozzi2003,Bozzi2005,deFlorian2011} in the case of the gluon fusion process. The matched \nlo{}+\nnll{} transverse momentum distribution can be calculated with the publicly available program {\tt HqT}~\cite{Bozzi2003,Bozzi2005,deFlorian2011}.

In this paper we show results for the transverse momentum spectrum of the Higgs boson in the gluon fusion process with exact top- and bottom-mass dependence, including the resummation of large logarithmic terms for small transverse momenta at \lo{}+\nll{}. We compare it to the effective theory approach and study the quality of the heavy-top limit to describe the exact top-mass cross section. Furthermore, we quantify the impact of the bottom-loop contribution in the \sm{}. We argue that the bottom-mass effects should be included at the order where their calculation is feasible and not by using a naive reweighting procedure in the heavy-top limit. As a result, we find that the effective theory approach approximates the exact top-mass dependence at the level of $0.5 \%$ in the region where resummation is important and to better than $4.5\%$ for transverse momenta below $150$\,GeV. The uncertainty induced by a missing bottom-quark contribution is of the order of $10\%$. Finally, we compare our results for the analytically resummed transverse momentum distribution to the ones obtained with a parton shower using the {\tt POWHEG} method in \reference{Bagnaschi2012} and find significant differences for small transverse momenta when considering both top- and bottom-mass effects.

The remainder of the paper is organized as follows: In the next section we will introduce the formalism we used for the transverse momentum resummation, indicate the resummation coefficients for Higgs production in gluon fusion and discuss the inclusion of the top- and bottom-mass effects in the resummed cross section. We outline our calculation in \sct{sec:outline}. In the final part of the paper we discuss the mass effects on three quantities: After recalling the known results for the total cross section and the \lo{} transverse momentum distribution, we present the resummed transverse momentum distribution of the Higgs including the full mass dependence at \lo{}+\nll{}. We conclude in \sct{sec:conclusions}.

\section{Transverse momentum resummation}\label{sec:resummation}
\subsection{Method}
\label{sec:resmethod}
In this section we sketch the formalism developed in \reference{Bozzi2005} to resum large logarithmic contributions of the transverse momentum ($p_T$) of a colorless particle in the final state. We stick to the case of a Higgs which is produced in gluon fusion.\footnote{The formalism in \reference{Bozzi2005} is based on the all-order transverse momentum resummation method developed in \reference{Dokshitzer:1978,Parisi:1979,Curci:1979,Collins:1981,Kodaira:1981,Davies:1984,Altarelli:1984,Collins:1984}.} We refer to \reference{Bozzi2005} and references therein for a more detailed description.

Due to collinear and soft singularities a fixed order transverse momentum distribution diverges at small $p_T$. This divergence is evident in the logarithmic $p_T$ structure of the fixed order (f.o.) cross section
\bal
\label{eq:fo}
\frac{d\sigma^{\text{f.o.}}}{dp_T^2} \sim \Bigg[\frac{\alpha_s}{\pi}\,\left( \frac{X^{(1:2)}}{p_T^2}\, \ln\left(\frac{m_H^2}{p_T^2}\right) +\frac{X^{(1;1)}}{p_T^2} + X^{(1;0)} + \mathcal{O}(p_T^2/m_H^2)\right)
%+\left(\frac{\alpha_s}{\pi}\right)^2 \bigg(X^{(1:4)}\, l_{p_T}^3 +X^{(1:3)}\, l_{p_T}^2 +X^{(1:2)}\, l_{p_T} +X^{(1:1)} + \mathcal{O}(p_T^2/m_H^2)\bigg)
+\mathcal{O}(\alpha_s^2)\Bigg],
\eal
where $m_H$ and $\alpha_s$ denote the Higgs mass and the strong coupling constant, respectively. The $X^{(1;a)}$ $(a=1,2)$ are introduced as the logarithmic coefficients and $X^{(1;0)}$ as the constant term in $p_T^2$ at \lo{}. To obtain a reliable cross section prediction for small transverse momenta, these logarithms have to be resummed to all orders in perturbation theory, leading to a finite distribution. In the notation of \reference{Catani:2011} the hadronic formula for the resummed logarithms integrated over the rapidity of the Higgs ($y$) reads\footnote{Note that we omit the coefficient $G$ introduced in \reference{Catani:2010} here and in what follows, since it enters at \nlo{}+\nnll{} which is beyond the accuracy needed in this paper. Furthermore, there is a subtle importance of different arguments of $\alpha_s$ in the original formula which is not expressed in this formula, since it is not essential for the purpose of this paper.}
\bal
\begin{split}
\label{eq:res}
\frac{d\sigma^{\text{res}}}{dp_T^2} = &\frac{m_H^2}{s}\,\sigma^{(0)}\,\int_{y_{\text{min}}}^{y_{\text{max}}}dy\int_0^\infty db\,\frac{b}2\,J_0(b\,p_T)\,S(m_H^2,b)\\&\sum\limits_{a,b=\{q,\bar{q},g\}}\,\int_{x_1}^1\frac{dz_1}{z_1}\int_{x_2}^1\frac{dz_2}{z_2}\,H\,C_{ga}(z_1)\,C_{gb}(z_2)\,f_a\left(\frac{x_1}{z_1},\frac{b_0^2}{b^2}\right)\,f_b \left(\frac{x_2}{z_2},\frac{b_0^2}{b^2}\right),
\end{split}
\eal
where the Sudakov factor is given by
\bal
\label{eq:sud}
S(m_H^2,b) = \exp\left\{-\int_{b_0^2/b^2}^{m_H^2}\,\frac{dq^2}{q^2}\left[A\,\ln\left(\frac{m_H^2}{q^2}\right)+B\right]\right\}.
\eal
The coefficients $A$, $B$, $C$ and $H$ are defined as a power series in $\alpha_s$
\bal
X&=\sum_{i=1}^{\infty}\left(\frac{\alpha_s}{\pi}\right)^i\, X^{(i)},\quad X\in\{A,B\}\\
C_{ab}(z)&=\delta_{ab}\,\delta(1-z)+\sum_{i=1}^{\infty}\left(\frac{\alpha_s}{\pi}\right)^i\, C^{(i)}_{ab}(z)\\
H&=1+\sum_{i=1}^{\infty}\left(\frac{\alpha_s}{\pi}\right)^i\, H^{(i)}.
\eal
$A$, $B$ and $C$ are process independent, but depend on the initial states of the \lo{} process ($gg$ or $q\bar{q}$). Accordingly, as soon as these coefficients are determined for one process, they are determined for all processes induced by the same initial states. The only  dependence on the process is embodied in the coefficient $H$, once a particular resummation scheme is choosen \cite{Catani:2000}.\footnote{Note that only when fixing the resummation scheme the coefficients $H$, $B$ and $C$ are unambiguously defined, since they are connected through so-called resummation-scheme transformations. Fixing $H$ (or $C$) for a single process amounts to fixing the resummation scheme \cite{Catani:2000}.} Instead of $H$ one can also determine the so-called partonic hard-collinear function $\mathcal{H}$ \cite{Catani:2011}, introduced in \reference{Bozzi2005}, which is defined through $C$ and $H$
\bal
\mathcal{H}_{gg\leftarrow ab}(z) &=H\,\int_0^1 dz_1\int_0^1 dz_2\,\delta(z-z_1z_2)\,C_{ga}(z_1)\,C_{gb}(z_2),
\eal
which again can be written as a power series in $\alpha_s$
\bal
\mathcal{H}_{gg\leftarrow ab}(z)&=\delta_{ga}\,\delta_{gb}\,\delta(1-z)+\sum_{i=1}^{\infty}\left(\frac{\alpha_s}{\pi}\right)^i\, \mathcal{H}^{(i)}_{gg\leftarrow ab}(z).
\eal
Let us make a few more comments about the resummation formula in eq. \eqref{eq:res}:
\begin{itemize}
\item The transverse momentum resummation has to be carried out in the impact parameter space ($b$-space) instead of $p_T$. The integration over $b$ is basically an inverse Fourier transform from $b$- to $p_T$-space. $J_0(b\,p_T)$ is the 0th-order Bessel function and the numerical coefficient $b_0$ is the Euler number $b_0=\gamma_E=0.577...$ .
\item $\sigma^{(0)}$ denotes the \lo{} cross section of the partonic process $gg\rightarrow H$.\footnote{Since the \lo{} of the process $gg\rightarrow H$ does not correspond to the \lo{} of the $p_T$ distribution of the Higgs, we will refer to $\sigma^{(0)}$ as the Born factor in the following to avoid confusion.}
\item The integration limits of the parton fractions $z_1$ and $z_2$ are $x_1=e^y\,m_H/\sqrt{s}$ and $x_2=e^{-y}\,m_H/\sqrt{s}$, where $s$ denotes the hadronic center-of-mass energy.
\item The functions $f_i$ denote the parton distribution functions (\pdf{}s).
\item For practical reasons (see, e.g., \reference{Bozzi2005}) transverse momentum resummation is usually performed in Mellin space ($N$-space). For this purpose one has to use the $N$-moments\footnote{The $N$-moments of a function $g(z)$ are defined as $g_N=\int_0^1\, dz\,z^{N-1}\,g(z)$.} of all quantities with respect to $z=m_H^2/\hat{s}$, where $\hat s$ denotes the partonic center-of-mass energy.
\item $d\sigma^{\text{res}}/dp_T^2$ in eq. \eqref{eq:res} fulfills the unitarity constraint, which is imposed in eq. (8) of \reference{Bozzi2005}. More precisely the coefficient $\mathcal{H}$ can be chosen in such a way that the integral over $p_T$ of the matched cross section, defined below, reproduces the total cross section. This serves as an important consistency check of the matched cross section.
\end{itemize}

Let us assume that the corresponding resummation coefficients of a process are known. Then we have to match the fixed order cross section in eq.~\eqref{eq:fo}, which is valid for high transverse momenta, and the resummed logarithmically enhanced contributions in eq.~\eqref{eq:res}, which describe the cross section in the small-$p_T$ region, to obtain a continuous transverse momentum distribution. For this purpose we subtract the logarithmic terms truncated at a fixed order
\bal
\label{eq:logs}
\frac{d\sigma^{\text{logs}}}{dp_T^2} = \left[\frac{d\sigma^{\text{res}}}{dp_T^2}\right]_{\text{f.o.}}
\eal
from the cross section in eq.~\eqref{eq:fo} at the same order. After that, we add back the logarithms resummed to all orders at the corresponding logarithmic accuracy (l.a.).
\bal
\label{eq:sig}
\left(\frac{d\sigma}{dp_T^2}\right)^{\text{f.o.}+\text{l.a.}} = \frac{d\sigma^{\text{f.o.}}}{dp_T^2}-\frac{d\sigma^{\text{logs}}}{dp_T^2} +\left[\frac{d\sigma^{\text{res}}}{dp_T^2}\right]_{\text{l.a.}}
\eal
This features two properties: First, the divergence of the fixed order cross section is canceled, leading to a finite result on the right hand side of eq.~\eqref{eq:sig}. Furthermore, double counting of logarithmic contributions present in both $d\sigma^{\text{f.o.}}/dp_T^2$ and $d\sigma^{\text{res}}/dp_T^2$ is avoided.

In the formalism of \reference{Bozzi2005} a new scale is introduced, namely the resummation scale $Q_{\text{res}}$. For simplicity we refrained from including it into the formulas of this section and set it equal to the Higgs mass. This scale may serve as an indicator of the uncertainty induced by the truncation of the cross section at some logarithmic accuracy. We will use it together with the other unphysical scales, namely the factorization ($\mu_F$) and renormalization scale ($\mu_R$), to obtain an error estimate of the cross section.

\subsection{Gluon fusion at \lo{}+\nll{}}\label{sec:resgf}
In this paper we consider transverse momentum resummation of the gluon fusion process at \lo{}+\nll{}, where the Higgs is produced through both a top- and a bottom-quark loop. While higher order corrections are usually evaluated in the heavy-top limit, our calculation includes the full dependence on the top ($m_t$) and bottom mass ($m_b$).

To calculate the resummed cross section in eq.~\eqref{eq:sig} one has to know the resummation coefficients described in \sct{sec:resmethod}. It is sufficient to determine the coefficient $A^{(1)}$ to resum all logarithms at leading logarithmic accuracy. For the full \nll{} resummation $A^{(1)}$, $A^{(2)}$, $B^{(1)}$ and $\mathcal{H}^{(1)}$ are needed, while at \nnll{} $A^{(3)}$, $B^{(2)}$ and $\mathcal{H}^{(2)}$ are required in addition.\footnote{See \reference{Bozzi2005}.}

For Higgs production in gluon fusion the \nll{} coefficients are\footnote{We set $\mu_F=\mu_R=Q_{\text{res}}=m_H$ throughout this chapter.}
\bal
A^{(1)}&= C_A,\\
A^{(2)}&= \frac12\,C_A\,\left[\left(\frac{67}{18}-\frac{\pi^2}6\right)-\frac59\,n_f\right],\\
B^{(1)}&= -\beta_0 = -\left(\frac{11}6\,C_A-\frac13\,n_f\right),\\
\mathcal{H}^{(1)}_{gg\leftarrow gg}(z)&= \delta(1-z)\left(C_A\,\frac{\pi^2}6+\frac12\,\mathcal{A}\right),\label{eq:H}\\
\mathcal{H}^{(1)}_{gg\leftarrow gq}(z) &= \mathcal{H}^{(1)}_{gg\leftarrow qg}(z)=-\frac12\,\hat P_{gq}^{\epsilon}(z)=\frac12\,C_F\,z,\\
\mathcal{H}^{(1)}_{gg\leftarrow q\bar{q}}(z) &= 0,
\eal
where $n_f=5$ is the number of light quark flavors, $C_A=3$ is a constant and $\mathcal{A}$ denotes the finite part of the virtual corrections\footnote{We define the finite part of the virtual according to eq.~(38) of \reference{deFlorian:2001}.}, which makes $\mathcal{H}^{(1)}_{gg\leftarrow gg}(z)$ process dependent. 

\subsection{Mass effects}\label{sec:me}
One of the main goals of this paper is the comparison of the resummed cross section with exact top and bottom masses to the heavy-top limit. Let us illustrate how the mass effects enter the cross section in eq.~\eqref{eq:sig}. 

As obvious from eq.~\eqref{eq:res} and \eqref{eq:logs} both $d\sigma^{\text{res}}/dp_T^2$ and $d\sigma^{\text{logs}}/dp_T^2$ are proportional to the Born factor $\sigma^{(0)}$. Thus these contributions calculated in the pure heavy-top limit can be reweighted by simply replacing the Born factor calculated with an infinite top mass $\sigma^{(0)}_{\text{htl}}$ by the Born factor with the full top-mass dependence $\sigma^{(0)}_{\text{t}}$ or with the full top- and bottom-mass dependence $\sigma^{(0)}_{\text{t+b}}$. After reweighting, there is no difference between those contributions in the heavy-top approach and those including the full mass dependence except for  the resummation coefficient $\mathcal{H}^{(1)}$ of the $gg$-channel, where the mass effects enter through the finite part of the virtual corrections $\mathcal{A}$, see eq.~\eqref{eq:H}.

All other mass effects concern the fixed order contribution $d\sigma^{\text{f.o.}}/dp_T$ of the resummed cross section, which enter already at the amplitude level. Nevertheless, in the limit $p_T\rightarrow 0$ the fixed order cross section factorizes as well into the Born factor $\sigma^{(0)}$ and a process independent splitting function. As a consequence, at fixed order the reweighted heavy-top limit and the cross section with full mass dependence have the same small-$p_T$ limit. This is essential and actually a mandatory condition, since the mass effects in $d\sigma^{\text{logs}}/dp_T^2$ factorize entirely into $\sigma^{(0)}$, see above, and it has to subtract the divergence of the fixed order cross section for small transverse momenta.

Since it will be needed throughout this paper, let us define the reweighted cross section as follows:
\bal
d\sigma_{\text{X}\rightarrow \text{Y}}= d\sigma_{\text{X}}\cdot\frac{\sigma_{\text{Y}}^{(0)}}{\sigma_{\text{X}}^{(0)}}.
\eal
E.g. $\sigma_{\text{htl}\rightarrow \text{t+b}} = \sigma_{\text{htl}}\cdot \left(\sigma_{\text{t+b}}^{(0)}/\sigma_{\text{htl}}^{(0)}\right)$ is the total cross section in the heavy-top limit reweighted by the Born factor including the full top- and bottom-mass dependence. In particular $d\sigma_{\text{htl}\rightarrow\text{t+b}}/dp_T$ is the transverse momentum distribution of $\sigma_{\text{htl}\rightarrow \text{t+b}}$. When the exact mass effects are not known, the reweighted cross section serves as a possible approximation of them. While the top-quark effects can usually be well described by the reweighted cross section in the heavy-top limit, it is clear that in general this is not true for the bottom-quark contribution. We will study this statement quantitatively in \sct{sec:results}.

\section{Outline of the calculation}
\label{sec:outline}
\begin{figure}[tb]
  \begin{center}
    \begin{tabular}{ccc}
      \mbox{\includegraphics[height=.12\textheight]{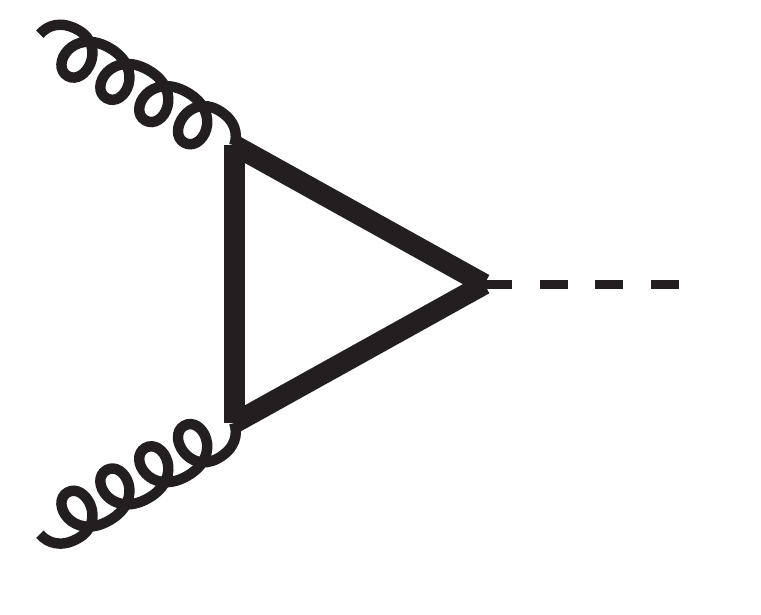}} & \hspace*{1.5cm}&
      \mbox{\includegraphics[height=.12\textheight]{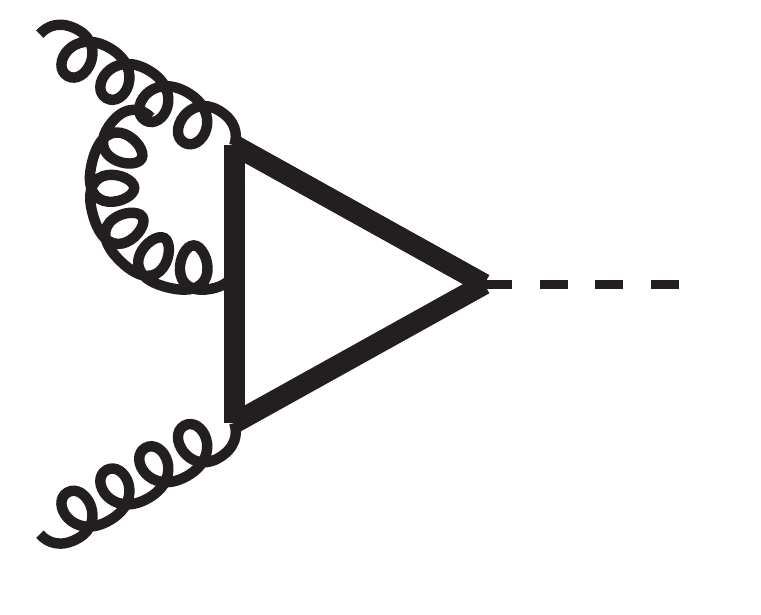}}
            \\
      (a) & & (b)
    \end{tabular}
    \parbox{.9\textwidth}{%
      \caption[]{\label{fig:lovirt}
        (a) \lo{} Feynman diagram of the gluon fusion process and (b) sample Feynman diagram of the virtual corrections. The graphical notation for the lines is: Thick straight $\hat=$ top or bottom quark;
        spiraled $\hat=$ gluon; dashed $\hat=$ Higgs boson.
        }
      }
  \end{center}
\end{figure}

Considering Higgs production in gluon fusion only the diagram shown in \fig{fig:lovirt}\,(a) has to be taken into account at \lo{}. The top quark gives the dominant contribution, since the Yukawa coupling is proportional to the corresponding quark mass. While the bottom loop still has a considerably large impact on the cross section, all other quarks can be safely omitted.

\begin{figure} [t]
  \begin{center}
    \begin{tabular}{ccc}
      \mbox{\includegraphics[height=.12\textheight]{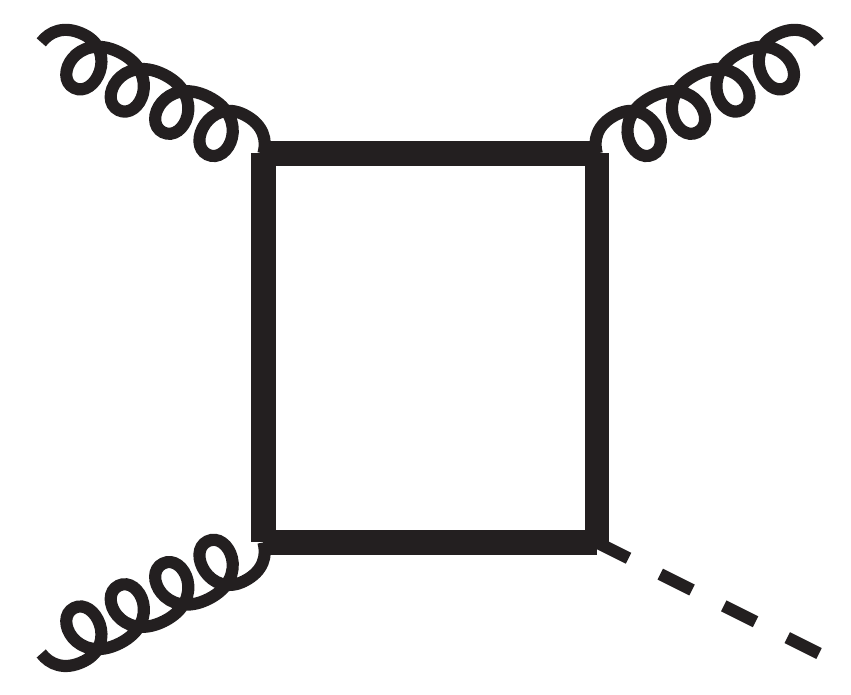}} &
      \mbox{\includegraphics[height=.12\textheight]{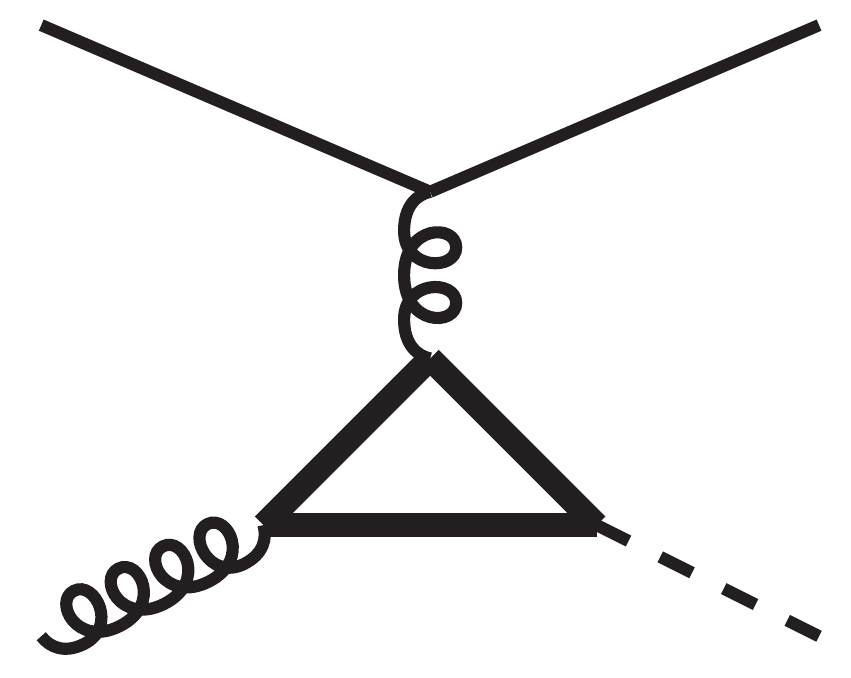}} &
      \raisebox{.7em}{\includegraphics[height=.12\textheight]{%
          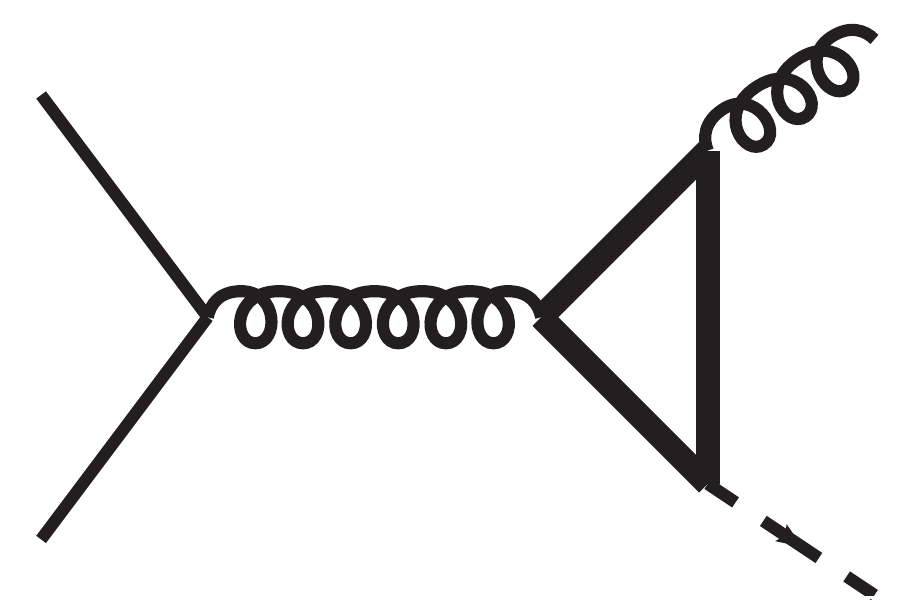}}
      \\
      (a) & (b) & (c)
    \end{tabular}
    \parbox{.9\textwidth}{
      \caption[]{\label{fig:real} Sample Feynman diagrams of the real corrections for the individual channels (a) $gg$ (b) $gq$ and (c) $q\bar{q}$, contributing to the process $pp\to H$ at \nlo{} \qcd{}. Notation as in
      \fig{fig:lovirt} and thin straight $\hat=$ light quark $q\in\{u,d,c,s,b\}$.
} }
  \end{center}
\end{figure}

At \nlo{} the virtual corrections, see \fig{fig:lovirt}\,(b), and the real emission diagrams in \fig{fig:real} have to be included. To avoid the approximation of an infinitely heavy top quark, all contributions need to be calculated by keeping the full top- and bottom-mass dependence. The total cross section has already been evaluated with exact top and bottom masses at \nlo{} some time ago \cite{Spira:1995}, see also \reference{Harlander:2005,Aglietti:2006,Anastasiou:2006,Muhlleitner:2006,Bonciani2007}.

In \fig{fig:lovirt} the Higgs' transverse momentum is always zero, since it is the only particle in the final state. Therefore, only the real emission diagrams in \fig{fig:real}, which contribute to the total cross section at \nlo{}, enter the transverse momentum distribution at \lo{} for $p_T>0$. Accordingly, these diagrams contribute to the first term on the right hand side of the resummed cross section in eq.~\eqref{eq:sig} at \lo{}+\nll{}:
\bal
\label{eq:siglo}
\left(\frac{d\sigma}{dp_T^2}\right)^{\text{\lo{}}+\text{\nll{}}} = \frac{d\sigma^{\text{\lo{}}}}{dp_T^2}-\left[\frac{d\sigma^{\text{res}}}{dp_T^2}\right]_{\text{\lo{}}} +\left[\frac{d\sigma^{\text{res}}}{dp_T^2}\right]_{\text{\nll{}}}.
\eal
As described in \sct{sec:resummation}, there are two more contributions to the resummed cross section that contain mass effects: The Born factor, which enters both in the second and third term on the right hand side of eq.~\eqref{eq:siglo}; and the finite part of the virtual corrections, which enters in the coefficient $\mathcal{H}^{(1)}$ of the $gg$-channel, see eq.~\eqref{eq:H}. These contributions with full mass dependence, the Born factor, the virtual and the real matrix elements, were obtained from the authors of \reference{Harlander:2010w}.\footnote{Those ingredients are known for a long time \cite{Georgi:1977,Ellis:1987,Spira:1995}.} We implemented them into a numerical program that calculates the resummed transverse momentum distribution at \lo{}+\nll{} analogous to {\tt HqT}~\cite{Bozzi2003,Bozzi2005,deFlorian2011}. The difference is that {\tt HqT} covers only the heavy-top limit, but up to \nlo{}+\nnll{}. The third term on the right hand side of eq.~\eqref{eq:siglo}, which resums the large logarithmic contributions, was obtained using a modified version of {\tt HqT}.

The resummation can be done separately for the individual sub-channels $gg$ (\fig{fig:real}\,(a)), $gq$ (\fig{fig:real}\,(b)) and $q\bar{q}$ (\fig{fig:real}\,(c)). The $gg$-channel contains double and single logarithmic contributions, while the $gq$-channel only embodies a single logarithmic divergence at small transverse momenta. The $q\bar{q}$-channel on the other hands remains finite for small $p_T$ on its own, since it contains no logarithmically enhanced terms at \lo{}. Consequently no resummation has to be performed in this channel.

Since the resummation procedure of \reference{Bozzi2005} fulfills the unitarity constraint\footnote{See \sct{sec:resmethod}.}, we were able to use this as an important cross-check of our calculation. For the integral of the resummed cross section over all $p_T$, we found agreement at the sub-percentage level with the total cross section \cite{Spira:1995}, separately for the individual sub-channels. We also checked that this integral is resummation scale independent. Furthermore, we compared the fixed order transverse momentum distribution to \reference{Harlander:2010w} and checked all contributions in the heavy-top limit against {\tt HqT}. In both cases we found perfect agreement.

\section{Results}\label{sec:results}
\subsection{Total cross section}\label{sec:total}
Before we discuss the transverse momentum distribution with exact top- and bottom-mass dependence, let us first summarize the quark-mass effects in case of the total inclusive cross section.

In the heavy-top limit, the total inclusive cross section has been calculated at \nnlo{} \cite{Harlander2002,Anastasiou2002,Ravindran2003}, while  the full quark-mass dependence is only known through \nlo{} \cite{Spira:1995}. One observes that the difference between the reweighted heavy-top limit and the exact top-mass dependence at \nlo{} is at the sub-percentage level for a Higgs mass lower than twice the top mass.  In particular this means that the corresponding ratios of the \nlo{} and \lo{} cross sections $K_{\text{htl}} = \sigma^{\nlo{}}_{\text{htl}} / \sigma^{\lo{}}_{\text{htl}}$ and $K_t = \sigma^{\nlo{}}_{\text{t}} / \sigma^{\lo{}}_{\text{t}}$ ($K$-factors) agree within this accuracy. A few years ago, finite top-mass effects have been investigated also at \nnlo{} \cite{Harlander2009,Harlander2010,Pak2009}, showing that these effects remain below $1\%$.

When the bottom-quark contribution is included, the total inclusive \nlo{} cross section is reduced by about $7\%$ for a Higgs mass of $125$\,GeV at the \lhc{} with $8$ TeV machine energy. At \lo{}, however, the cross section is lowered by more than $10\%$. Evidently, the \lo{} bottom-quark contribution is negative while the effect on the \nlo{} correction is positive. Hence the $K$-factors $K_{\text{htl}}$ and $K_{\text{t}}$ differ from $K_{\text{t+b}}$ at the order of $3.5\%$.

If the bottom contribution was not known at \nlo{}, one could try to approximate it through reweighting of $\sigma_{\text{t+b}}^{\lo{}}$ by $K_t$ or $K_{\text{htl}}$. A comparison with the exact \nlo{} cross section $\sigma^{\nlo{}}_{\text{t+b}}$ shows that it is better to not reweight the \nlo{} cross section and therefore to omit the \nlo{} bottom-quark contribution completely, but take into account only the exact \lo{} $\mbottom$ dependence.
 
A similar conclusion might be true for differential quantities as well. Therefore we present the full quark-mass dependence of the resummed transverse momentum distribution in this paper and investigate the mass effects.

\subsection{Transverse momentum distribution}\label{sec:pt}

\subsubsection{Preliminary remarks}
After some considerations concerning the $p_T$ distribution at \lo{}, we will show the resummed cross section with respect to the transverse momentum of the Higgs at \lo{}+\nll{} including the full top- and bottom-mass dependence. We present results for the \lhc{} at $8$ TeV and a Higgs mass of $125$\,GeV. We use the \mstw{} \nlo{} \pdf{} sets \cite{Martin2009} and the corresponding input of the strong coupling constant $\alpha_s(m_Z)=0.12018$, where $m_Z$ denotes the $Z$-Boson mass. We insert an on-shell top and bottom mass of $m_t = 172$\,GeV and $m_b = 4.9$\,GeV, respectively. Our central scale choice is $\mu_0=m_H/2$ for the renormalization, factorization and the resummation scale.%\footnote{Please note that in the formalism used in this paper only logarithms of the form $\ln(m_H/p_T)$ are resummed.  For the bottom contribution also logarithms of the form $\ln(m_b/m_H)$ occur.  Their resummation, though desirable, is beyond the scope of the present paper.}.

Let us assess at this point the validity of the resummation approach for the bottom contribution, whose dominant contribution arises from the interference terms of diagrams with top and bottom loop. The resummation formalism is well established in case of the heavy-top approximation \cite{Bozzi2005} introduced in \sct{sec:resmethod}. In this case there is a clear hierarchy between the scale $p_T$ where the resummation procedure is valid and the scales of the process, namely $m_H$ and $m_t$. Evidently, the same is true when including the full top-mass dependence. However, in case of the bottom loop the situation is different. The logarithms arising within the calculation are of the form $\ln(m_H^2/p_T^2)$ and $\ln(m_b^2/p_T^2)$. Though we correctly resum all logarithms in $p_T$, arriving at a finite cross section, setting $Q_{\text{res}}\sim m_H$ does not keep track of the scale that accompanies $p_T$ inside the logarithms. Consequently, logarithms of the form $\ln(m_b^2/m_H^2)$ are not properly resummed within our calculation. The separate resummation of logarithms in $m_b^2/p_T^2$ and $m_H^2/p_T^2$, though desirable, is beyond the scope of the present paper. 
Furthermore, since the dominant contribution is given by the interference terms between top and bottom loop, choosing a resummation scale $Q_{\text{res}}\sim m_H$ should serve a good approximation. Nevertheless we will discuss the comparison of the resummed cross section for $Q_{\text{res}}\sim m_b$ and $Q_{\text{res}}\sim m_H$ in the final part of the manuscript to give the reader complementary information about their difference.

\label{sec:locons}
\begin{figure}[t]
  \begin{center}
    \begin{tabular}{c}
      \mbox{\includegraphics[height=.4\textheight]{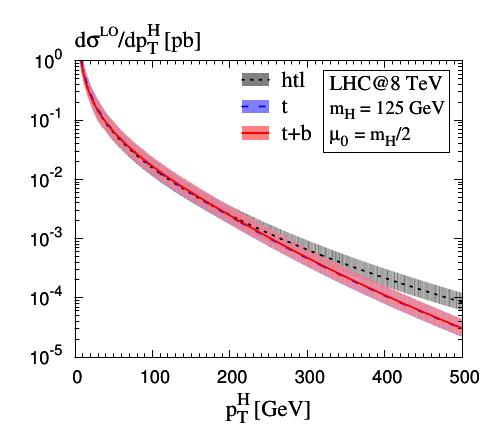}}
    \end{tabular}
    \parbox{.9\textwidth}{
      \caption[]{\label{fig:LOpT} \lo{} transverse momentum distribution of the Higgs in the heavy-top limit (black, dotted), including the exact top mass (blue, dashed) and with exact top- and bottom-mass dependence (red, solid). The uncertainty bands are obtained by scale variation. In particular we varied one scale within $[0.5\,\mu_0,2\,\mu_0]$, while fixing the other, and vice versa. The maximum and minimum values of the cross section in this procedure are the uncertainties to the central value. In this logarithmic plot the curve of $d\sigma_{\text{t}}/dp_T$ is almost indistinguishable from $d\sigma_{\text{t+b}}/dp_T$.}
      }
  \end{center}
\end{figure}

\subsubsection{\lo{} considerations}
Before analyzing the quark-mass effects on the resummed transverse momentum distribution at \lo{}+\nll{}, let us recall the situation for the pure \lo{} $p_T$ distribution. It is shown in \fig{fig:LOpT} for the heavy-top limit $d\sigma_{\text{htl}}^{\lo{}}/dp_T$ (black, dotted), exact top-mass dependence $d\sigma_t^{\lo{}}/dp_T$ (blue, dashed) and including both top- and bottom-quark masses $d\sigma_{\text{t+b}}^{\lo{}}/dp_T$ (red, solid). As expected, the fixed order curves diverge in the limit $p_T \rightarrow 0$. Furthermore, all three curves are quite close to each other for small transverse momenta, while for $p_T > 200$\,GeV a gap between the heavy-top limit and the other two curves emerges. Since the top contribution is dominant, $d\sigma_t^{\lo{}}/dp_T$ and $d\sigma_{t+b}^{\lo{}}/dp_T$ stay close to each other as well for high transverse momenta. The uncertainty bands are obtained through independent variation of $\mu_R$ and $\mu_F$ within $[0.5\,\mu_0,2\,\mu_0]$.

\begin{figure}[bht]
  \begin{center}
    \begin{tabular}{c}
      \mbox{\includegraphics[height=.4\textheight]{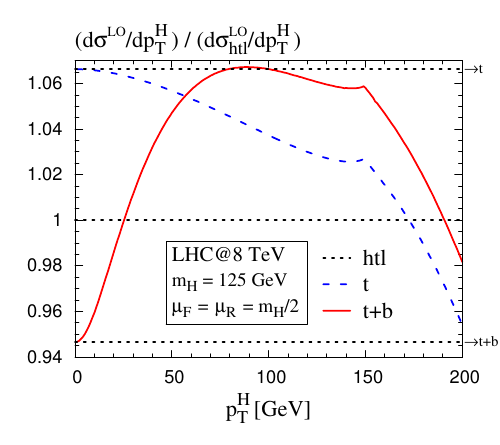}}
    \end{tabular}
    \parbox{.9\textwidth}{
      \caption[]{\label{fig:LOrel} Curves of \fig{fig:LOpT} normalized to the heavy-top limit curve. The upper and lower black dotted curves denote the reweighted cross sections $d\sigma^{\lo{}}_{\text{htl}\rightarrow t}/dp_T$ and $d\sigma^{\lo{}}_{\text{htl}\rightarrow \text{t+b}}/dp_T$, respectively.}
      }
  \end{center}
\end{figure}
Considering the relative contribution of the mass effects normalized to the heavy-top limit, see \fig{fig:LOrel}, we can visualize discrepancies also for small transverse momenta. For comparison we show three different curves for the heavy-top limit (black, dotted): The pure heavy-top limit is the normalization in this plot and hence located at one. The other two lines denote the reweighted cross sections $d\sigma^{\lo{}}_{\text{htl}\rightarrow \text{t}}/dp_T$ (upper line) and $d\sigma^{\lo{}}_{\text{htl}\rightarrow \text{t+b}}/dp_T$ (lower line), respectively.

Let us first compare the cross section with full top-mass dependence (blue, dashed) to the reweighted cross section $d\sigma^{\lo{}}_{\text{htl}\rightarrow t}/dp_T$. We find that, similar to what was found in \reference{Harlander2012}, the reweighted heavy-top limit works for the top loop to better than $4\%$ as long as $p_T < 150$\,GeV. For large transverse momenta the heavy-top approximation is not valid any more, because the discrepancy is growing rapidly for $p_T > 150$\,GeV. The kink at $p_T \approx 150$\,GeV occurs in the $q\bar{q}$ channel and is due to the kinematical cut at $\sqrt{\hat{s}} = \sqrt{p_T^2+m_H^2}+p_T$. Because of the factorization of the cross section in the limit $p_T\rightarrow 0$, see \sct{sec:me}, $d\sigma^{\lo{}}_{\text{t}}/dp_T$ and $d\sigma^{\lo{}}_{\text{htl}\rightarrow \text{t}}/dp_T$ in \fig{fig:LOrel} become identical in this limit.

The same is true for the cross section including top- and bottom-mass dependence (red, solid) in \fig{fig:LOrel}: $d\sigma^{\lo{}}_{\text{t+b}}/dp_T$ converges to $d\sigma^{\lo{}}_{\text{htl}\rightarrow \text{t+b}}/dp_T$ in the limit $p_T\rightarrow 0$. Let us compare the cross section with exact top and bottom masses to the pure heavy-top limit (black, dotted) located at one. The discrepancy ranges between $-5\%$ and $+7\%$ for $p_T < 200$\,GeV. The difference is even higher (up to $13\%$) when we compare $d\sigma^{\lo{}}_{\text{t+b}}/dp_T$ to the reweighted cross section $d\sigma^{\lo{}}_{\text{htl}\rightarrow \text{t+b}}/dp_T$.  Evidently, as soon as the bottom loop is considered, omitting the mass effects works better than approximating them by the reweighted cross section $d\sigma^{\lo{}}_{\text{htl}\rightarrow \text{t+b}}/dp_T$ . This is similar to the total cross section, where the \nlo{} $K$-factors $K_{\text{htl}}$ and $K_{\text{t}}$ should not be applied to the bottom-quark contribution. In the following we will investigate if a similar conclusion can be drawn for the resummed transverse momentum distribution.

\begin{figure}[bht]
  \begin{center}
    \begin{tabular}{cc}
      \mbox{\includegraphics[height=.32\textheight]{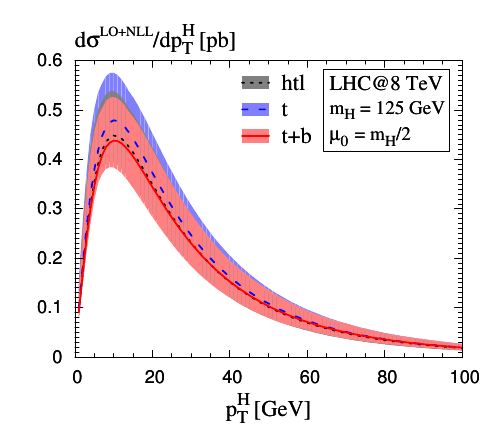}} &
      \mbox{\includegraphics[height=.32\textheight]{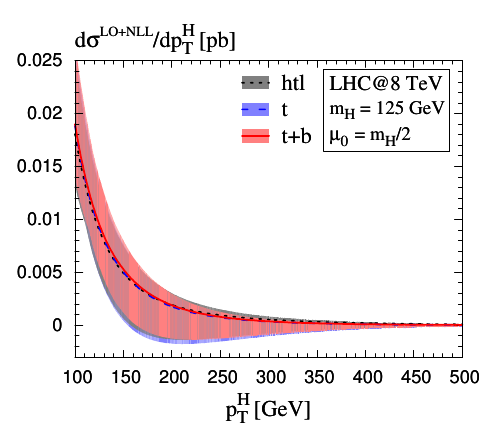}}\\
      (a) &  (b)
    \end{tabular}
    \parbox{.9\textwidth}{
      \caption[]{\label{fig:res} Same as \fig{fig:LOpT}, but for the resummed cross sections and splitted into two ranges: (a) $0$--$100$\,GeV and (b) $100$--$500$\,GeV.}
      }
  \end{center}
\end{figure}
\subsubsection{Resummed cross section at \lo{}+\nll{}}\label{sec:resummed}
\fig{fig:res} shows the matched cross sections of the fixed order prediction and resummed large logarithmic contributions at \lo{}+\nll{}, defined in eq.~\eqref{eq:siglo}. As before we plot three different curves, the pure heavy-top approximation (black, dotted), the cross section including full top-mass dependence (blue, dashed) and the one with exact top- and bottom-quark masses (red, solid). Unlike the \lo{} distribution, the resummed cross sections are finite at small transverse momenta, leading to a reliable prediction also for $p_T \lesssim 30$\,GeV.

The uncertainties of the cross sections emerge from the truncation of perturbative series with respect to both the strong coupling constant and the logarithmic accuracy. They are estimated by the variation of the renormalization and factorization scale and the resummation scale, respectively. The error bands are obtained through independent variation of $\mu_R$, $\mu_F$ and $Q_{\text{res}}$ within $[0.5\,\mu_0,2\,\mu_0]$. We neglect the uncertainty emerging from the \pdf{}s and $\alpha_s$. It was already found for the heavy-top limit, see \reference{Bozzi2005} Fig.~8, that for $Q_{\text{res}}\gtrsim m_H$ the matched cross section can become negative for high transverse momenta. Accordingly, the uncertainty bands of all cross sections in \fig{fig:res} become negative around $p_T \sim 150$\,GeV because of the resummation scale variation. For $p_T \lesssim 100$\,GeV the relative scale uncertainty ranges between $10\%$ and $50\%$, while in average they are of the order of $25\%$. For higher transverse momenta the uncertainty gets huge, due to the fact that the error bands become negative.

\begin{figure}[bht]
  \begin{center}
    \begin{tabular}{c}
      \mbox{\includegraphics[height=.4\textheight]{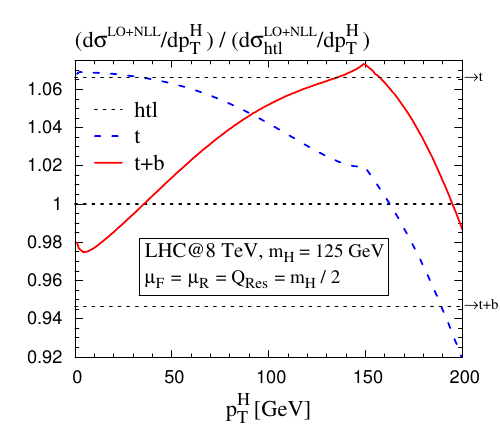}}
    \end{tabular}
    \parbox{.9\textwidth}{
      \caption[]{\label{fig:resrel} Same as \fig{fig:LOrel} but for the resummed cross sections.}
      }
  \end{center}
\end{figure}

Let us investigate the mass effects on the resummed cross section in the low $p_T$ range in more detail. For this purpose the relative contributions normalized to the heavy-top limit are plotted in \fig{fig:resrel}. Just as in \fig{fig:LOrel} there are three curves for the heavy-top limit (black, dotted). The one in the middle denotes the pure heavy-top limit, while the other two denote the reweighted cross sections. The upper line represents $d\sigma^{\lo{}+\nll{}}_{\text{htl}\rightarrow t}/dp_T$ and the lower one $d\sigma^{\lo{}+\nll{}}_{\text{htl}\rightarrow \text{t+b}}/dp_T$.

Let us first compare the cross section with full top-mass dependence (blue, dashed) to the reweighted cross section $d\sigma^{\lo{}+\nll{}}_{\text{htl}\rightarrow \text{t}}/dp_T$. We find that the reweighted heavy-top limit is working to better than $4.5\%$ as long as $p_T < 150$\,GeV, while, as expected, the discrepancy grows as the transverse momentum of the Higgs increases. In the region $p_T < 50$\,GeV, where resummation effects are important, the top-mass effects remain even below $0.5 \%$. In the small-$p_T$ limit the resummed cross section including the exact top-quark mass is not identical to $d\sigma^{\lo{}+\nll{}}_{\text{htl}\rightarrow t}/dp_T$ in \fig{fig:resrel} as it was for the \lo{} distribution. This is caused by the fact that the mass effects do not completely factorize into the Born factor $\sigma^{(0)}_t$, see \sct{sec:me}.

The same is true for the resummed cross section with full top- and bottom-mass dependence (red, solid) in \fig{fig:resrel}, where the gap to $d\sigma^{\lo{}+\nll{}}_{\text{htl}\rightarrow \text{t+b}}/dp_T$ in the limit $p_T \rightarrow 0$ is even larger. Let us investigate the mass effects when both top and bottom quark are included and determine the quality of the reweighted cross section $d\sigma^{\lo{}+\nll{}}_{\text{htl}\rightarrow \text{t+b}}/dp_T$ to approximate them. Therefore, we compare the red, solid curve ($d\sigma^{\lo{}+\nll{}}_{\text{t+b}}/dp_T$) to the lower black, dotted line ($d\sigma^{\lo{}+\nll{}}_{\text{htl}\rightarrow \text{t+b}}/dp_T$) and find that the discrepancy reaches up to $14\%$ for $p_T < 200$\,GeV. In fact omitting the mass effects completely works better: The difference between 
$d\sigma^{\lo{}+\nll{}}_{\text{t+b}}/dp_T$ and the pure heavy-top limit ranges only from $-2.5\%$ to $7\%$ for $p_T<200$\,GeV. Note also that omitting the $b$-loop contribution and account only for the top-mass effects through $d\sigma^{\lo{}+\nll{}}_{\text{htl}\rightarrow \text{t}}/dp_T$ is closer to $d\sigma^{\lo{}+\nll{}}_{\text{t+b}}/dp_T$ than $d\sigma^{\lo{}+\nll{}}_{\text{htl}\rightarrow \text{t+b}}/dp_T$, since in this case the difference reaches only up to $9\%$. 

\begin{figure}[t]
  \begin{center}
    \begin{tabular}{c}
      \mbox{\includegraphics[height=.4\textheight]{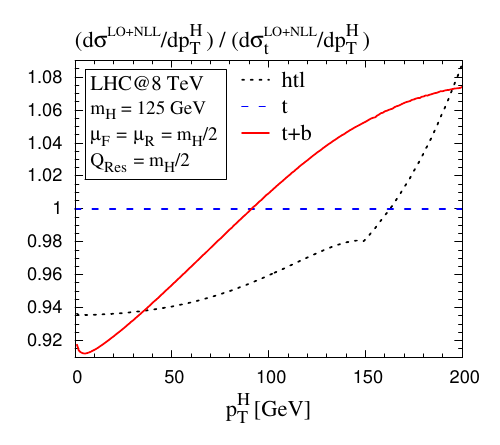}}
    \end{tabular}
    \parbox{.9\textwidth}{
      \caption[]{\label{fig:topbot} Resummed cross sections normalized to the one including the full top-mass effects.}
      }
  \end{center}
\end{figure}

Regarding this, it is interesting to examine the bottom-mass effects assuming the exact top-mass dependence is known. In \fig{fig:topbot} the ratio of the resummed cross section including top- and bottom-mass dependence and the one including only top-mass effects is shown (red, solid). Omitting the $b$-loop effects amounts to about $\pm 8 \%$ when the top mass is taken into account exactly.

\begin{figure}[thb]
  \begin{center}
    \begin{tabular}{c}
      \mbox{\includegraphics[height=.4\textheight]{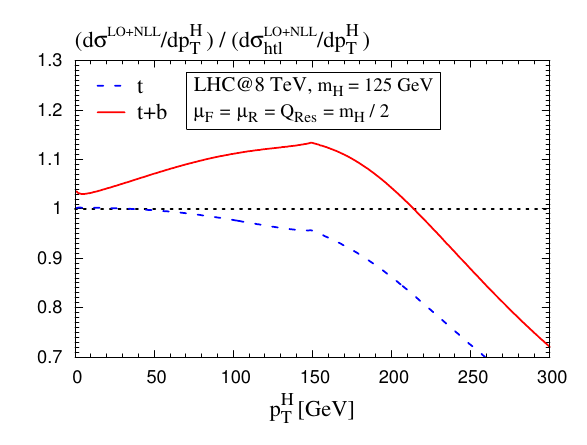}}
    \end{tabular}
    \parbox{.9\textwidth}{
      \caption[]{\label{fig:comp} Analogous to lower right plot in Fig.~3 of \reference{Bagnaschi2012}. Blue, dashed curve: Ratio of $d\sigma_{\text{t}}/dp_T$ and $d\sigma_{\text{htl}\rightarrow\text{t}}/dp_T$; red, solid curve: Ratio of $d\sigma_{\text{t+b}}/dp_T$ and $d\sigma_{\text{htl}\rightarrow\text{t+b}}/dp_T$.}
      }
  \end{center}
\end{figure}
In an earlier study \cite{Bagnaschi2012} the mass effects on the resummed transverse momentum distribution were calculated using the {\tt POWHEG} method~\cite{Alioli:2008} in combination with the {\tt PYTHIA} parton shower \cite{Sjostrand2006}. For comparison we reproduced the lower right plot in Fig.~3 of \reference{Bagnaschi2012} in the case of the analytically resummed cross section at \lo{}+\nll{}, see \fig{fig:comp}. It shows $d\sigma_{\text{t}}/dp_T$ normalized to $d\sigma_{\text{htl}\rightarrow \text{t}}/dp_T$ (blue, dashed) and $d\sigma_{\text{t+b}}/dp_T$ normalized to $d\sigma_{\text{htl}\rightarrow \text{t+b}}/dp_T$ (red, solid). Although the parameter choices of \reference{Bagnaschi2012} are slightly different, for $p_T \lesssim 200$\,GeV the top-mass effects on the analytically resummed cross section are in good agreement with the ones of the resummed cross section obtained with {\tt POWHEG}+{\tt PYTHIA} (blue, dashed in \fig{fig:comp}; black, dashed in Fig.~3 of \reference{Bagnaschi2012}). Considering an exact top- and bottom-mass dependence (red, solid curve in our plot; blue, solid curve in \reference{Bagnaschi2012}), on the other hand, the mass effects on the two approaches appear to be considerably different for $p_T\lesssim 50$ GeV, i.e. in the region where resummation becomes important. However, both approaches are theoretically well defined and the numerical results are consistent within the respective resummation formalism. The discrepancy might be caused by a different treatment of logarithms in $m_b/m_H$ in the two approaches. Their treatment in the present paper is discussed in the introduction of \sct{sec:pt}. Furthermore, it needs to be clarified whether the discrepancy arises from the normalization factor, or whether it is a genuine effect in the cross section with full mass dependence. Currently, it has to be considered as a measure of the theory uncertainty at small $p_T$. Clearly, the source of the difference deserves further investigation.\footnote{According to \reference{Frixione} in the MC@NLO approach \cite{Frixione2002} the shape of the curve including top- and bottom-mass dependence is much more similar to ours (red, solid curve in \fig{fig:comp}).}

In summary we find that in all cases studied in this paper it is not a good approximation to account for $b$-loop effects using the reweighted cross section in the heavy-top limit. We conclude that bottom-mass effects should be included only up to the order where their calculation is feasible and should be omitted otherwise.

Along these lines we give a recommendation for the evaluation of the best prediction regarding the resummed cross section of the gluon fusion process. As stated above, transverse momentum resummation of the Higgs boson is known in the heavy-top limit at \nlo{}+\nnll{} \cite{Bozzi2003,Bozzi2005,deFlorian2011}, while in this paper we presented the full top- and bottom-mass dependence at \lo{}+\nll{}. Both cross sections should be combined by
\begin{align}
\label{best}
\frac{d\sigma_{\text{best}}}{dp_T} = \frac{d\sigma_{\text{htl}\rightarrow\text{t}}^{\nlo{}+\nnll{}}}{dp_T}-\frac{d\sigma_{\text{htl}\rightarrow\text{t}}^{\lo{}+\nll{}}}{dp_T}+\frac{d\sigma_{\text{t+b}}^{\lo{}+\nll{}}}{dp_T}.
\end{align}
The first two terms on the right hand side of eq.~\eqref{best} can be calculated with the program {\tt HqT}~\cite{Bozzi2003,Bozzi2005,deFlorian2011}. As well for the second but especially for the third term numbers can be obtained from the authors of this paper upon request.

\begin{figure}[bht]
  \begin{center}
    \begin{tabular}{cc}
      \hspace*{-0.5cm}\mbox{\includegraphics[height=.32\textheight]{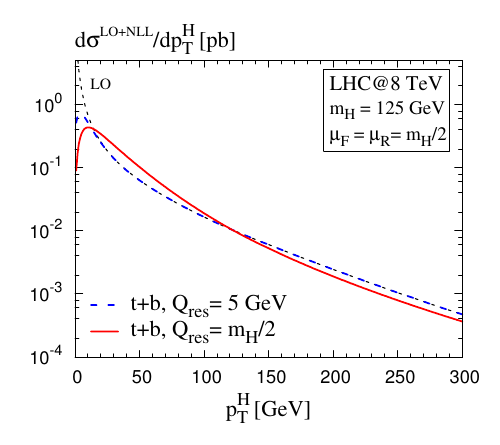}} &
      \hspace*{-0.6cm}\mbox{\includegraphics[height=.32\textheight]{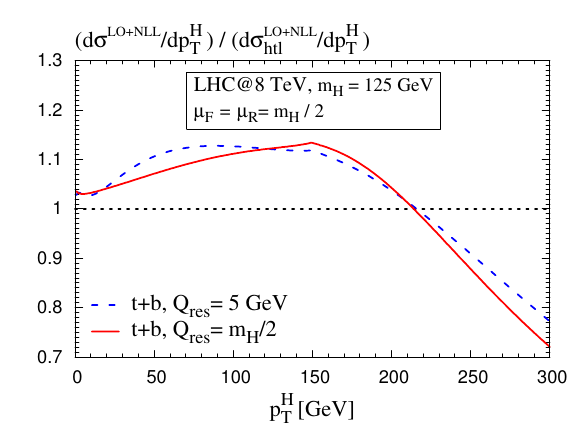}}\\
      (a) &  (b)
    \end{tabular}
    \parbox{.9\textwidth}{
      \caption[]{\label{fig:resscale} (a) Resummed cross section for $Q_{\text{res}}=m_H/2$ (red, solid) and $Q_{\text{res}}= 5$ GeV (blue, dashed); the \lo{} curve is shown for comparison. (b) Same as \fig{fig:comp}, but with $Q_{\text{res}}$ from (a). The red, solid curve is identical to \fig{fig:comp}.}
      }
  \end{center}
\end{figure}

Let us return at this point to the question of choosing an appropriate resummation scale for the bottom contribution. For this purpose \fig{fig:resscale}~(a) compares the resummed cross section with top- and bottom-mass dependence for $Q_{\text{res}}=m_H/2$ (red, solid) and $Q_{\text{res}}=5$ GeV\footnote{Please recall that such a scale choice is not suitable for the top contribution in general. Consequently, the purpose of this comparison is just to provide qualitative information about the scale of the bottom contribution.} (blue, dashed). For a low resummation scale the impact of resummed logarithms is small at high transverse momenta. Thus already for $p_T>25$ GeV the resummed cross section with $Q_{\text{res}}=5$ GeV differs from the \lo{} cross section (black, dotted) at the sub-percentage level. This leads to the big differences in absolute numbers observed in \fig{fig:resscale}~(a) between the red, solid and blue, dashed line.

However, if we consider the relative top- and bottom-mass effects, see \fig{fig:resscale}~(b), the resulting curves are rather similar. Each cross section in \fig{fig:resscale}~(b) is normalized to its corresponding cross section in the heavy-top limit using the same resummation scale. As before, the curve for $Q_{\text{res}}=m_H/2$ is red, solid and the one for $Q_{\text{res}}=5$ GeV is blue, dashed. The overall behaviour of the curves is nearly the same. The main difference is the scope of the curves in the region $p_T<150$ GeV.  The red, solid curve has a linear behaviour when going from $150$ GeV to vanishing $p_T$, while the blue, dashed curve has the same plateau between $50$ and $150$ GeV as the red, solid \lo{} curve in \fig{fig:LOrel}. This behaviour originates again from the fact that for a low resummation scale the logarithms only affect small transverse momenta.

Since there are only slight differences between the relative curves, such a curve is well suited to be used for reweighting. The resulting cross section then hardly depends on the corresponding resummation scale. We suggest to reweight the heavy-top limit calculated at some resummation scale $Q$ according to

\begin{align}
\frac{d\sigma}{dp_T} = \frac{d\sigma_{\text{htl}}^{\nlo{}+\nnll{}}(Q)}{dp_T}\cdot \left(\frac{d\sigma_{\text{t+b}}^{\lo{}+\nll{}}(Q')/dp_T}{d\sigma_{\text{htl}}^{\lo{}+\nll{}}(Q')/dp_T}\right),
\end{align}

while the term in the brackets is calculated at some scale $Q'$. Then the result has a rather mild dependence on $Q'$.

\section{Conclusions}\label{sec:conclusions}
We presented the transverse momentum resummation of the Higgs boson in gluon fusion including the full top- and bottom-mass dependence at \lo{}+\nll{}. We found that top-mass effects can be approximated in the heavy-top limit to better than $0.5 \%$ for $p_T < 50$\,GeV and $4.5\%$ for $p_T < 150$\,GeV, respectively. Furthermore, we analyzed the contribution of the $b$-loop in the resummed cross section. It amounts to about $10 \%$. In comparison with an earlier study, we found that for small transverse momenta the influence of the bottom mass on the analytically resummed transverse momentum distribution is considerably different from the one obtained using the {\tt POWHEG} method in combination with the {\tt PYTHIA} parton shower, while the top-mass effects on both approaches are in good agreement. We also showed that the bottom-mass effects are not well approximated by the reweighted cross section in the heavy-top limit. As a consequence we recommended a best prediction for the resummed cross section that combines two calculations done with current technology: The \lo{}+\nll{} contribution should be subtracted from the \nlo{}+\nnll{} prediction in the heavy-top limit and added back including the full top- and bottom-mass dependence.

\paragraph{Acknowledgements.}
We would like to thank Robert Harlander and Anurag Tripathi for fruitful discussion and enlightening comments, and the authors of \reference{Bagnaschi2012} and Stefano Frixione for helpful communication. This work was supported by {\abbrev BMBF} contracts 05H09PXE and 05H12PXE,
and the Helmholtz Alliance ``Physics at the Terascale''.

\bibliographystyle{elsarticle-mod}
\bibliography{gghpt}

\begin{thebibliography}{10}
\expandafter\ifx\csname url\endcsname\relax
  \def\url#1{\texttt{#1}}\fi
\expandafter\ifx\csname urlprefix\endcsname\relax\def\urlprefix{URL }\fi
\expandafter\ifx\csname href\endcsname\relax
  \def\href#1#2{#2} \def\path#1{#1}\fi

\bibitem{Collaboration2012}
{CMS Collaboration}, Phys. Lett. B 716 (2012)  30--61.

\bibitem{ATLAS2012}
{ATLAS Collaboration}, Phys. Lett. B 716 (2012)  1--29.

\bibitem{Dittmaier2011}
S.~Dittmaier~et al., arXiv:1101.0593   , arXiv:1201.3084.

\bibitem{Harlander2002}
R.~V. Harlander, W.~B. Kilgore, Phys. Rev. Lett. 88 (2002)  201801.

\bibitem{Anastasiou2002}
C.~Anastasiou, K.~Melnikov, Nucl. Phys. B 646 (2002)  220--256.

\bibitem{Ravindran2003}
V.~Ravindran, J.~Smith, W.~van Neerven, Nucl. Phys. B 665 (2003)  325--366.

\bibitem{Marzani2008}
S.~Marzani, R.~D. Ball, V.~{Del Duca}, S.~Forte, A.~Vicini, Nucl. Phys. B 800
  (2008)  127--145.

\bibitem{Harlander2009}
R.~V. Harlander, K.~J. Ozeren, JHEP 0911 (2009)  088.

\bibitem{Harlander2010}
R.~V. Harlander, H.~Mantler, S.~Marzani, K.~J. Ozeren, Eur. Phys. J. C 66
  (2010)  359--372.

\bibitem{Pak2009}
A.~Pak, M.~Rogal, M.~Steinhauser, Phys. Lett. B 679 (2009)  473--477, JHEP 1002
  (2010) 025.

\bibitem{Harlander2012}
R.~V. Harlander, T.~Neumann, K.~J. Ozeren, M.~Wiesemann, JHEP 1208 (2012)  139.

\bibitem{DelDuca2001a}
V.~{Del Duca}, W.~Kilgore, C.~Oleari, C.~Schmidt, D.~Zeppenfeld, Nucl. Phys. B
  616 (2001)  367--399.

\bibitem{Alwall2012}
J.~Alwall, Q.~Li, F.~Maltoni, Phys. Rev. D 85 (2012)  014031.

\bibitem{Bagnaschi2012}
E.~Bagnaschi, G.~Degrassi, P.~Slavich, A.~Vicini, JHEP 1202 (2012)  088.

\bibitem{Harlander:2003}
R.~Harlander, Eur. Phys. J. C 33 (2004)  S454--S456.

\bibitem{deFlorian:2009}
D.~de~Florian, M.~Grazzini, Phys. Lett. B 674 (2009)  291--294.

\bibitem{Anastasiou:2010}
C.~Anastasiou, R.~Boughezal, F.~Petriello, JHEP 1006 (2010)  101.

\bibitem{Baglio2010}
J.~Baglio, A.~Djouadi, JHEP 1010 (2010)  064, JHEP 1103 (2011) 055.

\bibitem{Spira:1995}
M.~Spira, A.~Djouadi, D.~Graudenz, R.~Zerwas, Nucl. Phys. B 453 (1995)  17--82.

\bibitem{Alioli:2008}
S.~Alioli, P.~Nason, C.~Oleari, E.~Re, JHEP 0904 (2009)  002, JHEP 1006 (2010)
  043.

\bibitem{deFlorian:1999}
D.~de~Florian, M.~Grazzini, Z.~Kunszt, Phys. Rev. Lett. 82 (1999)  5209--5212.

\bibitem{Ravindran:2002dc}
V.~Ravindran, J.~Smith, W.~van Neerven, Nucl. Phys. B 634 (2002)  247--290.

\bibitem{Glosser:2002}
C.~J. Glosser, C.~R. Schmidt, JHEP 0212 (2002)  016.

\bibitem{Anastasiou2004}
C.~Anastasiou, K.~Melnikov, F.~Petriello, Phys. Rev. Lett. 93 (2004)  262002.

\bibitem{Catani2007}
S.~Catani, M.~Grazzini, Phys. Rev. Lett. 98 (2007)  222002.

\bibitem{catani:1988}
S.~Catani, E.~D'Emilio, L.~Trentadue, Phys. Lett. B 211 (1988)  335--342.

\bibitem{Yuan:1991}
C.-P. Yuan, Phys. Lett. B 283 (1992)  395--402.

\bibitem{kauffman:1991}
R.~Kauffman, Phys. Rev. D 45 (1992)  1512--1517.

\bibitem{Bozzi2003}
G.~Bozzi, S.~Catani, D.~de~Florian, M.~Grazzini, Phys. Lett. B 564 (2003)
  65--72.

\bibitem{Bozzi2005}
G.~Bozzi, S.~Catani, D.~de~Florian, M.~Grazzini, Nucl. Phys. B 737 (2006)
  73--120.

\bibitem{deFlorian2011}
D.~Florian, G.~Ferrera, M.~Grazzini, D.~Tommasini, JHEP 1111 (2011)  064, JHEP
  1206 (2012) 132.

\bibitem{Dokshitzer:1978}
Y.~Dokshitzer, D.~Dyakonov, S.~Troyan, Phys. Rep. 58 (1980)  269--395.

\bibitem{Parisi:1979}
G.~Parisi, R.~Petronzio, Nucl. Phys. B 154 (1979)  427--440.

\bibitem{Curci:1979}
G.~Curci, M.~Greco, Y.~Srivastava, Nucl. Phys. B 159 (1979)  451--468.

\bibitem{Collins:1981}
J.~C. Collins, D.~E. Soper, Nucl. Phys. B 193 (1981)  381--443, Nucl. Phys. B
  197 (1982) 446--476.

\bibitem{Kodaira:1981}
J.~Kodaira, L.~Trentadue, Phys. Lett. B 112 (1982)  66--70.

\bibitem{Davies:1984}
C.~Davies, W.~Stirling, Nucl. Phys. B 244 (1984)  337--348.

\bibitem{Altarelli:1984}
G.~Altarelli, R.~Ellis, M.~Greco, G.~Martinelli, Nucl. Phys. B 246 (1984)
  12--44.

\bibitem{Collins:1984}
J.~Collins, D.~E. Soper, G.~Sterman, Nucl. Phys. B 250 (1985)  199--224.

\bibitem{Catani:2011}
S.~Catani, M.~Grazzini, Eur. Phys. J. C 72 (2012)  2013.

\bibitem{Catani:2010}
S.~Catani, M.~Grazzini, Nucl. Phys. B 845 (2011)  297--323.

\bibitem{Catani:2000}
S.~Catani, D.~de~Florian, M.~Grazzini, Nucl. Phys. B 596 (2001)  299--312.

\bibitem{deFlorian:2001}
D.~de~Florian, M.~Grazzini, Nucl. Phys. B 616 (2001)  247--285.

\bibitem{Harlander:2005}
R.~V. Harlander, P.~Kant, JHEP 0512 (2005)  015.

\bibitem{Aglietti:2006}
U.~Aglietti, R.~Bonciani, G.~Degrassi, A.~Vicini, JHEP 0701 (2007)  021.

\bibitem{Anastasiou:2006}
C.~Anastasiou, S.~Beerli, S.~Bucherer, A.~Daleo, Z.~Kunszt, JHEP 0701 (2007)
  082.

\bibitem{Muhlleitner:2006}
M.~M\"{u}hlleitner, M.~Spira, Nucl. Phys. B 790 (2008)  1--27.

\bibitem{Bonciani2007}
R.~Bonciani, G.~Degrassi, A.~Vicini, JHEP 0711 (2007)  095.

\bibitem{Harlander:2010w}
R.~V. Harlander, F.~Hofmann, H.~Mantler, JHEP 1102 (2011)  055.

\bibitem{Georgi:1977}
H.~M. Georgi, S.~L. Glashow, M.~E. Machacek, D.~V. Nanopoulos, Phys. Rev. Lett.
  40 (1978)  692--694.

\bibitem{Ellis:1987}
R.~Ellis, I.~Hinchliffe, M.~Soldate, J.~{Van Der Bij}, Nucl. Phys. B 297 (1988)
   221--243.

\bibitem{Martin2009}
A.~D. Martin, W.~J. Stirling, R.~S. Thorne, G.~Watt, Eur. Phys. J. C 63 (2009)
  189--285.

\bibitem{Sjostrand2006}
T.~Sj\"{o}strand, S.~Mrenna, P.~Skands, JHEP 0605 (2006)  026, Comput. Phys.
  Commun. 178 (2008) 852--867.

\bibitem{Frixione}
S.~Frixione, private   communication.

\bibitem{Frixione2002}
S.~Frixione, B.~R. Webber, JHEP 0206 (2002)  029.

\end{thebibliography}

\end{document}